%% file: semi-inclusive.tex
\documentclass[a4paper,12pt]{article}
\usepackage[left=20mm,right=20mm,top=20mm,bottom=30mm]{geometry}
\usepackage{amsmath}
\usepackage{amssymb}
\usepackage[affil-it]{authblk}
\usepackage{hyperref}
\usepackage{microtype}
\usepackage{setspace}
\usepackage{tikz}
\usepackage{cite}

\input{tikz}

\hypersetup{
  colorlinks,
  linkcolor={blue!50!black},
  citecolor={blue!50!black},
  urlcolor={blue!50!black}
}

\newcommand{\abs}[1]{\lvert #1 \rvert}

\newcommand{\tikzvector}[6]{
  \draw [->]
    ([xshift=#3, yshift=#4]$#1!0.25!#2$)
    to node [#5] {#6}
    ([xshift=#3, yshift=#4]$#1!0.75!#2$)
}

\DeclareMathOperator{\Tr}{Tr}

\begin{document}

\title{$pp$ scattering at the LHC with the lepton pair production and one
  proton tagging}

\author[1]{S.~I.~Godunov}
\author[1]{E.~K.~Karkaryan}
\author[1]{V.~A.~Novikov}
\author[2]{A.~N.~Rozanov}
\author[1]{M.~I.~Vysotsky}
\author[1]{E.~V.~Zhemchugov}

\affil[1]{
  \small I.~E.~Tamm Department of Theoretical Physics, Lebedev Physical
  Institute, \newline
  53 Leninskiy Prospekt, Moscow, 119991, Russia
}
\affil[2]{
  \small Centre de Physique de Particules de Marseille (CPPM), Aux-Marseille
  Universite, CNRS/IN2P3, \newline
  163 avenue de Luminy, case 902, Marseille, 13288, France
}

\date{}

\maketitle

\begin{abstract}
  Analytical formulas for the cross section of the reaction $pp \to p + \ell^+
  \ell^- + X$ are presented. Fiducial cross sections are compared with those
  measured recently by the ATLAS collaboration.
\end{abstract}

\section{Introduction}

Lepton pairs produced in ultraperipheral collisions (UPC) of protons at the
Large Hadron Collider (LHC) are accompanied by forward scattering of the
protons. Previous measurements of this process were performed by the ATLAS
collaboration without proton tagging~\cite{1708.04053}. We have calculated the
cross section for this reaction with the help of the equivalent photon
approximation (EPA) in~\cite{2106.14842}, taking into account the so-called
survival factor which addresses the diminishing of the cross section because of
proton disintegration due to strong interactions.  The result obtained agrees
with the measurement within the experimental accuracy.  Predictions for the
cross section were also made with the help of Monte-Carlo
calculations~\cite{0803.0883, 1512.01178, 1410.2983, 1508.02718}.

The CMS and TOTEM collaborations have reported statistically significant
proton-tagged dilepton production~\cite{1803.04496}, but the cross sections have
not been measured.  The ATLAS collaboration has managed to measure the cross
sections~\cite{2009.14537}. In the events selected for the analysis, one of the
scattered protons is detected by the ATLAS Forward Proton Spectrometer.
The other proton could remain intact, in which case it could or could not be
detected by the opposite forward detector, or it could disintegrate.  The
transversal momentum of the lepton pair $p_T^{\ell \ell}$ was required to be
less than 5~GeV. This momentum equals to the sum of transversal momenta of the
photons emitted by the protons. The transversal momentum of the photon that was
emitted by the proton that survived the collision cannot be much higher than
$\hat q = 0.2$~GeV~\cite{1806.07238}.  Therefore, the transversal momentum of
the second photon has to be less than 5~GeV.

In what follows we derive analytical formulas which describe the fiducial cross
sections measured in~\cite{2009.14537}. These formulas allow for simple
numerical integration instead of the usual Monte Carlo approach and thus can
provide intuitive insights into the process targeted by the experiment. Our
numerical results are in the ballpark of experimental data, while their
substantial deviation would signal New Physics. This paper is a continuation
of the work presented in our previous paper~\cite{2112.01870} where we have
derived formulas for the cross section without experimental cuts on the leptons
phase space and with no requirement for any of the protons to hit the forward
detectors.

In Section~\ref{s:exclusive} we calculate the fiducial cross section for the
case of elastic scattering of the second proton. The contribution to the
fiducial cross section of the processes when the second proton disintegrates is
calculated in Section~\ref{s:inclusive}. We conclude in
Section~\ref{s:conclusion}.

\section{Fiducial cross section for the reaction $pp \to p + \ell^+ \ell^- + p$}

\label{s:exclusive}

The spectrum of photons radiated by proton is~\cite{1806.07238}
\begin{equation}
  n_p(\omega)
  = \frac{2 \alpha}{\pi \omega}
    \int\limits_0^\infty \frac{D(Q^2)}{Q^4} q_\perp^3 \, \mathrm{d} q_\perp,
  \label{proton-spectrum}
\end{equation}
where $\alpha$ is the fine structure constant, $Q^2 \equiv -q^2 = q_\perp^2 +
\omega^2 / \gamma^2$, $q$ is the photon 4-momentum, $q_\perp$ is the photon
transverse momentum, $\omega$ is the photon energy, $\gamma = E_p / m_p \approx
6.93 \cdot 10^3$ is the Lorentz factor of the proton, $E_p = 6.5$~TeV is the
proton energy, $m_p$ is the proton mass,\footnote{
  In Ref.~\cite{1806.07238}, the Dirac form factor squared was used instead of
  $D(Q^2)$. It leads to incorrect accounting for the magnetic form factor.
}
\begin{equation}
  D(Q^2) = \frac{G_E^2(Q^2) + \frac{Q^2}{4 m_p^2} G_M^2(Q^2)}
                {1 + \frac{Q^2}{4 m_p^2}},
  \label{form-factor}
\end{equation}
$G_E(Q^2)$ and $G_M(Q^2)$ are the Sachs electric and magnetic form factors. For
the Sachs form factors we use the dipole approximation:
\begin{equation}
  G_E(Q^2) = \frac{1}{(1 + Q^2 / \Lambda^2)^2},
  ~
  G_M(Q^2) = \frac{\mu_p}{(1 + Q^2 / \Lambda^2)^2},
  ~
  \Lambda^2 = \frac{12}{r_p^2} = 0.66~\text{GeV}^2,
  \label{sachs-form-factors}
\end{equation}
where $\mu_p = 2.79$ is the proton magnetic moment, and $r_p = 0.84$~fm is the
proton charge radius~\cite{rmp93-025010}.
Substituting~\eqref{form-factor}, \eqref{sachs-form-factors}
into~\eqref{proton-spectrum}, we obtain the following analytical
expression:
\begin{equation}
  \begin{split}
    n_p(\omega)
    &= \frac{\alpha}{\pi \omega}
       \left\{
           \left( 1 + 4 u - (\mu_p^2 - 1) \frac{u}{v} \right)
           \ln \left( 1 + \frac{1}{u} \right)
         - \frac{24 u^2 + 42 u + 17}{6 (u+1)^2}
    \right. \\ & \left.
         - \frac{\mu_p^2 - 1}{(v - 1)^3}
           \left[
              \frac{1 + u / v}{v - 1}
              \ln \frac{u + v}{u + 1}
            - \frac{
                6u^2 (v^2 - 3v + 3) + 3u (3v^2 - 9v + 10) + 2v^2 - 7v + 11
              }{6 (u + 1)^2}
           \right]
       \right\},
  \end{split}
\end{equation}
where
\begin{equation}
  u = \left( \frac{\omega}{\Lambda \gamma} \right)^2, \ 
  v = \left( \frac{2 m_p}{\Lambda} \right)^2.
\end{equation}

The cross section for the lepton pair production in the case when both of the
protons survive is
\begin{equation}
  \sigma(pp \to p + \ell^+ \ell^- + p)
  = \int\limits_0^\infty \int\limits_0^\infty
    \sigma(\gamma \gamma \to \ell^+ \ell^-)
    \, n_p(\omega_1)
    \, n_p(\omega_2)
    \, \mathrm{d} \omega_1
    \, \mathrm{d} \omega_2,
  \label{proton-xsection}
\end{equation}
where $\sigma(\gamma \gamma \to \ell^+ \ell^-)$ is the cross section for the
production of a lepton pair in a collision of two real photons with energies
$\omega_1$ and $\omega_2$. The plus signs in the reaction notation indicate
large rapidity gaps between the produced particles.

In~\cite{2009.14537}, as well as in all other measurements made at the LHC, the
phase space is constrained by the following requirements: $p_{i,T} > \hat p_T$,
$\abs{\eta_i} < \hat \eta$, where $p_{i,T}$ is the transversal momentum of
lepton $\ell_i$, and $\eta_i$ is its pseudorapidity. In the case of muons $\hat
p_T = 15$~GeV and $\hat \eta = 2.4$, while in the case of electrons $\hat p_T =
18$~GeV and $\hat \eta = 2.47$. The transversal momentum of the lepton pair is
equal to the sum of the transversal momenta of the photons and is limited by
$\sqrt{Q^2} \lesssim \hat q = 0.2$~GeV, so transversal momenta of the leptons
are equal with good accuracy: $p_T \equiv p_{1,T} \approx p_{2,T}$.

Neglecting the lepton mass, the differential fiducial cross section in this case
is
\begin{equation}
  \frac{\mathrm{d} \sigma_\text{fid.}(pp \to p + \ell^+ \ell^- + p)}
       {\mathrm{d} W}
  = \int\limits_{\max \left( \hat p_T, \frac{W}{2 \cosh \hat \eta} \right)}
               ^{W / 2}
    \mathrm{d} p_T
    \, \frac{\mathrm{d} \sigma(\gamma \gamma \to \ell^+ \ell^-)}{\mathrm{d} p_T}
    \, \frac{\mathrm{d} \hat L}{\mathrm{d} W},
  \label{xsection}
\end{equation}
where $W = \sqrt{4 \omega_1 \omega_2}$ is the invariant mass of the lepton pair,
\begin{equation}
  \frac{\mathrm{d} \sigma(\gamma \gamma \to \ell^+ \ell^-)}
       {\mathrm{d} p_T}
  = \frac{8 \pi \alpha^2}{W^2 p_T}
    \cdot \frac{1 - 2 p_T^2 / W^2}{\sqrt{1 - 4 p_T^2 / W^2}}
  \label{xsection-pT}
\end{equation}
\cite{landau-lifshitz-4}, $\mathrm{d} \hat L / \mathrm{d} W$ is the
photon-photon luminosity taking into account the limits on the phase space,
\begin{equation}
  \frac{\mathrm{d} \hat L}{\mathrm{d} W}
  = \frac{W}{2}
    \int\limits_{-\hat y}^{\hat y}
       n_p \left( \tfrac{W}{2} \, \mathrm{e}^y \right)
    \, n_p \left( \tfrac{W}{2} \, \mathrm{e}^{-y} \right)
    \, \mathrm{d} y,
  \label{luminosity-no-fd}
\end{equation}
$y = \tfrac12 \ln \tfrac{\omega_1}{\omega_2}$ is the rapidity of the lepton
pair, and the integration over $\omega_1$ and $\omega_2$
in~\eqref{proton-xsection} is changed to the integration over $W$ and $y$:
$\mathrm{d} \omega_1 \mathrm{d} \omega_2 = \frac{W}{2} \mathrm{d} W \mathrm{d}
y$. The pseudorapidities of charged leptons for given values of $W$ and $p_T$
are determined by the value of $y$. In this way the cut on $\eta$ $(-\hat \eta <
\eta < \hat \eta)$ is transformed to the cut on $y$: $-\hat y < y < \hat y$,
where~\cite[(B.9)]{1806.07238}\footnote{Ref.~\cite{1806.07238} uses $x =
\omega_1 / \omega_2 \equiv \mathrm{e}^{2 y}$.}
\begin{equation}
  \hat y
  = \hat \eta
  + \frac{1}{2}
    \ln \frac{1 - \sqrt{1 - 4 p_T^2 / W^2}}{1 + \sqrt{1 - 4 p_T^2 / W^2}}.
\end{equation}
Here $\hat y$ has to be greater than zero; this requirement leads to the
inequality $p_T > \frac{W}{2 \cosh \hat \eta}$ in the lower limit of the
integration with respect to $p_T$ in~\eqref{xsection}.

Another requirement imposed in~\cite{2009.14537} is that one of the protons hits
the forward detector. To do that, the proton must lose a fraction of its energy
$\xi$: $\xi_\text{min} < \xi < \xi_\text{max}$, where $\xi_\text{min} = 0.035$,
$\xi_\text{max} = 0.08$. This translates to limits on the energy of the photon
emitted by this proton: 
\begin{equation}
  227~\text{GeV} \equiv \omega_\text{min}
  < \omega_1
  < \omega_\text{max} \equiv 520~\text{GeV}.
  \label{omega-limits}
\end{equation}
To take that into account, the integration limits in
eq.~\eqref{luminosity-no-fd} have to be narrowed:
\begin{equation}
  \frac{\mathrm{d} \hat L_\text{FD}}{\mathrm{d} W}
  = \frac{W}{2}
    \int\limits_{\max(-\hat y, \tilde y)}^{\min(\hat y, \tilde Y)}
       n_p \left( \tfrac{W}{2} \, \mathrm{e}^y \right)
    \, n_p \left( \tfrac{W}{2} \, \mathrm{e}^{-y} \right)
    \, \mathrm{d} y,
  \label{luminosity}
\end{equation}
where FD stands for the ``forward detector'',
\begin{equation}
  \begin{aligned}
    \tilde y &= \ln \max \left(
      \frac{2 \omega_{1, \text{min}}}{W}, \frac{W}{2 \omega_{2, \text{max}}}
    \right), \\
    \tilde Y &= \ln \min \left(
      \frac{2 \omega_{1, \text{max}}}{W}, \frac{W}{2 \omega_{2, \text{min}}}
    \right),
  \end{aligned}
\end{equation}
$\omega_{1, \text{min}}$, $\omega_{1, \text{max}}$, $\omega_{2, \text{min}}$,
$\omega_{2, \text{max}}$ are the limits on energy losses for each of the
protons.\footnote{
  This change may result in poor numerical convergence of the integral with
  respect to $p_T$ in~\eqref{xsection} when $\tilde y$ and $\tilde Y$ have the
  same sign. To address that, the lower integration limit in~\eqref{xsection}
  should be replaced with $\max \left( \hat p_T, \frac{W}{2 \cosh \hat \eta},
  \frac{W}{2 \cosh(\max(\tilde y, -\tilde Y) - \hat \eta)} \right)$.
}
Calculation with
\begin{equation}
  \begin{aligned}
       \omega_{1, \text{min}} &= \omega_\text{min},
    &  \omega_{2, \text{min}} &= 0,
    \\ \omega_{1, \text{max}} &= \omega_\text{max},
    &  \omega_{2, \text{max}} &= \infty
  \end{aligned}
  \label{ω-one}
\end{equation}
will result in the fiducial cross section with the first proton hitting the
forward detector. Calculation with
\begin{equation}
  \begin{aligned}
       \omega_{1, \text{min}} &= \omega_\text{min},
    &  \omega_{2, \text{min}} &= \omega_\text{min},
    \\ \omega_{1, \text{max}} &= \omega_\text{max},
    &  \omega_{2, \text{max}} &= \omega_\text{max}
  \end{aligned}
  \label{ω-both}
\end{equation}
yields the fiducial cross section with both protons hitting the forward
detectors.  To calculate the cross section measured in~\cite{2009.14537}, one
has to multiply the former by 2 and subtract the latter to avoid double
counting:
\begin{equation}
  \sigma_\text{fid,\cite{2009.14537}}(pp \to p + \ell^+ \ell^- + p)
  = 2 \sigma_\text{fid}(pp \to p + \ell^+ \ell^- + p) \rvert_\eqref{ω-one}
  - \sigma_\text{fid}(pp \to p + \ell^+ \ell^- + p) \rvert_\eqref{ω-both}.
  \label{double-counting}
\end{equation}

In both measurements in~\cite{2009.14537}, the selected region of invariant
masses of lepton pairs was $W > 20$~GeV with the region $70~\text{GeV} < W <
105~\text{GeV}$ excluded to suppress the background from $Z$ decays. Collecting
together all of the phase space constraints relevant to the exclusive process
(see Ref.~[94] in~\cite{2009.14537}), we get:
\begin{itemize}
  \item $20~\text{GeV} < W < 70~\text{GeV}$ or $W > 105$~GeV.
  \item $0.035 < \xi < 0.08$ which is equivalent to $227~\text{GeV} < \omega <
  520~\text{GeV}$.
  \item For muons:
  \begin{itemize}
    \item $\hat p_T = 15$~GeV, $\hat \eta = 2.4$.
    \item $\sigma_\text{fid,\cite{2009.14537}}(pp \to p + \mu^+ \mu^- + p) =
    8.6$~fb.
  \end{itemize}
  \item For electrons:
  \begin{itemize}
    \item $\hat p_T = 18$~GeV, $\hat \eta = 2.47$.
    \item $\sigma_\text{fid,\cite{2009.14537}}(pp \to p + e^+ e^- + p) =
    10.1$~fb.
  \end{itemize}
\end{itemize}

\section{Fiducial cross section for the reaction $pp \to p + \ell^+ \ell^- + X$}

\label{s:inclusive}

\begin{figure}
  \centering
  \begin{tikzpicture}
    \coordinate (Pin)  at (-4,  3);
    \coordinate (Pout) at ( 2,  3);
    \coordinate (Qin)  at (-4, -3);
    \coordinate (Qout) at ( 2, -3);
    \coordinate (GP)   at (-2,  3);
    \coordinate (GQ)   at (-2, -3);
    \coordinate (GL1)  at ( 0,  1);
    \coordinate (GL2)  at ( 0, -1);
    \coordinate (Lout) at ( 2,  1);
    \coordinate (Lin)  at ( 2, -1);

    \draw [fermion] (Pin) node [left] {$p$} -- (GP);
    \draw [fermion] (GP) -- (Pout) node [right] {$p$};
    \draw [fermion] (Qin) node [left] {$q$} -- (GQ);
    \draw [fermion] (GQ) -- (Qout) node [right] {$q$};
    \draw [fermion] (Lin) node [right] {$\ell$} -- (GL2);
    \draw [fermion] (GL2) -- (GL1);
    \draw [fermion] (GL1) -- (Lout) node [right] {$\ell$};
    \draw [photon]  (GP) -- node [above right] {$\gamma$} (GL1);
    \draw [photon]  (GQ) -- node [below right] {$\gamma$} (GL2);

    \tikzvector{(Pin)}{(GP)}{0pt}{-2mm}{midway, below}{$p_1$};
    \tikzvector{(0, 3)}{(Pout)}{0pt}{-2mm}{midway, below}{$p'_1$};
    \tikzvector{(Qin)}{(GQ)}{0pt}{2mm}{midway, above}{$p_2$};
    \tikzvector{(0, -3)}{(Qout)}{0pt}{2mm}{midway, above}{$p'_2$};
    \tikzvector{(GP)}{(GL1)}{-1.4mm}{-1.4mm}{midway, below left}{$q_1$};
    \tikzvector{(GQ)}{(GL2)}{-1.4mm}{1.4mm}{midway, above left}{$q_2$};
    \tikzvector{(GL1)}{(Lout)}{0pt}{2mm}{midway, above}{$k_1$};
    \tikzvector{(GL2)}{(Lin)}{0pt}{-2mm}{midway, below}{$k_2$};
  \end{tikzpicture}
  \caption{Lepton pair production in semiexclusive reaction}
  \label{f:semiexclusive}
\end{figure}
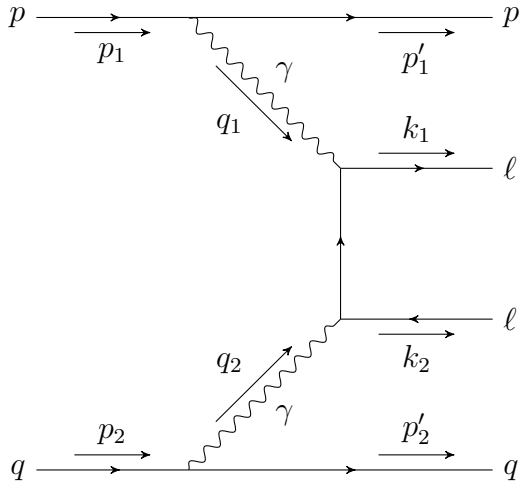

In this section we calculate the cross section for lepton pair production with
one of the protons scattered elastically and detected by the forward detector,
while the other proton disintegrates. Following the parton model, we consider
this process as a two-photon lepton pair production in a collision of a proton
and a quark, summed over all quarks:
\begin{equation}
  \sigma(pp \to p + \ell^+ \ell^- + X)
  = \sum\limits_q \sigma(pq \to p + \ell^+ \ell^- + q).
\end{equation}
One of the Feynman diagrams of the $p q \to p + \ell^+ \ell^- + q$ reaction is
presented in Fig.~\ref{f:semiexclusive}.
In the laboratory system we have the
following expressions for the momenta of the colliding particles:
\begin{equation}
  p_1 = (E, 0, 0, E), ~ p_2 = (xE, 0, 0, -xE),
  \label{p1-p2}
\end{equation}
where $x$ is the fraction of the momentum of the disintegrating proton carried
by the quark.  In the following we will integrate over $x$ from a value much
less than 1 to 1, but the numerically important values of $x$ are of the order
of $1/3$, so we have neglected here the masses of the proton $m_p$ and the quark
$m_q$: $E / m_p = \gamma \approx 6.93 \cdot 10^3 \gg 1$, $xE / m_q \gg 1$ even
for the (sea) $b$ quark.

The momentum of the photon emitted by the proton,
\begin{equation}
  q_1
  = \left(
      \omega_1, \vec q_{1 \perp}, \frac{\omega_1 \gamma}{\sqrt{\gamma^2 - 1}}
    \right),
  \label{q1}
\end{equation}
where $\vec q_{1 \perp}$ is its transversal momentum. The 4-momentum squared,
\begin{equation}
  Q_1^2 \equiv -q_1^2 \approx q_{1 \perp}^2 + (\omega_1 / \gamma)^2,
\end{equation}
is limited by the condition that the proton survives: $Q_1^2 \lesssim \hat
q^2 \approx (0.2~\text{GeV})^2$~\cite{1806.07238}, so the first photon is
approximately real: $Q_1^2 \ll W^2$. Therefore, the equivalent photon
approximation can be used here.

For the second photon, we also apply the equivalent photon approximation, but
with a correction.\footnote{Another option would be to use the parton
distribution for photons, see for example~\cite{1708.01256}.}  Since the quark
is bound within the proton, we use the constituent quark mass $m_q = m_p/3
\approx 300~\text{MeV}$ to describe its
properties.\footnote{\label{fn:quark-mass}Variation of $m_q$ from $200$ to
$400$~MeV changes the values of the cross sections presented in
Eqs.~\eqref{inclusive-mu}, \eqref{inclusive-e} by less than 1\%. It is clear
that at the present level of experimental accuracy the uncertainty due to the
choice of the quark mass is negligible.} The momentum of the second photon
\begin{equation}
  q_2
  \approx \left(
    \omega_2,
    \vec q_{2 \perp},
    -\frac{\omega_2 \gamma_q}{\sqrt{\gamma_q^2 - 1}}
  \right),
  \label{q2}
\end{equation}
and its square is
\begin{equation}
  Q_2^2 \equiv -q_2^2 \approx q_{2 \perp}^2 + (\omega_2 / \gamma_q)^2,
  \label{virtuality2}
\end{equation}
where $\gamma_q = E_q / m_q = 3 x E_p / m_p = 3 x \gamma$.  In the ATLAS
measurement the following upper bound was effectively imposed on $q_{2 \perp}$:
$\abs{\vec q_{1 \perp} + \vec q_{2 \perp}} \sim q_{2 \perp} < \hat p_T^{\ell
\ell} = 5$~GeV.

The square of the invariant mass of the lepton pair $W^2 = (q_1 + q_2)^2 \approx
2 q_1 q_2 + q_2^2 \approx 4 \omega_1 \omega_2 - Q_2^2$. For $W \sim 100$~GeV,
$Q_2^2 \ll W^2$, and $\omega_2 \approx W^2 / 4 \omega_1$. Here $\omega_1$ is
bounded by the requirement that the proton hits the forward
detector~\eqref{omega-limits}.  Therefore, $\omega_2$ is of the order of a few
GeV~--- the lepton pair will have large rapidity towards the detector that
registers the proton. We assume $\omega_1 \ll E$ and $\omega_2 \ll xE$ in the
following.

To derive the cross section for the $pq \to p + \ell^+ \ell^- + q$ reaction, we
follow closely Ref.~\cite{prep15-181} and begin with its Eq.~(5.1) (multiplied
by the parton distribution function $f_q(x, Q_2^2)$):
\begin{multline}
  \mathrm{d} \sigma(pq \to p + \ell^+ \ell^- + q)
  = \frac{Q_q^2 (4 \pi \alpha)^2}{q_1^2 q_2^2}
    \rho_1^{\mu \nu} \rho_2^{\alpha \beta}
    M_{\mu \alpha} M^*_{\nu \beta}
    \frac{(2 \pi)^4 \delta^{(4)}(q_1 + q_2 - k_1 - k_2) \mathrm{d} \Gamma}
         {4 \sqrt{(p_1 p_2)^2 - p_1^2 p_2^2}}
    \times \\
    \frac{\mathrm{d}^3 p'_1}{(2 \pi)^3 2 E'_1}
    \frac{\mathrm{d}^3 p'_2}{(2 \pi)^3 2 E'_2}
    \times f_q(x, Q_2^2) \mathrm{d} x,
  \label{master}
\end{multline}
where $Q_q$ is the electric charge of quark $q$, $\rho_1^{\mu \nu}$ and
$\rho_2^{\alpha \beta}$ are the photon density matrices for the photons emitted
by the proton or the quark respectively, $M_{\mu \alpha}$ is the amplitude for
the $\gamma^* \gamma^* \to \ell^+ \ell^-$ process, $\mathrm{d} \Gamma$ is the
phase space element of the lepton pair, $E'_1$ and $E'_2$ are the energies of
the proton and the quark after the collision.

Since $p'_i = p_i - q_i$, we rewrite the phase space element of the proton and
the quark as follows:
\begin{equation}
  \frac{\mathrm{d}^3 p'_1}{E'_1} \cdot \frac{\mathrm{d}^3 p'_2}{E'_2}
  \approx \frac{\mathrm{d}^2 q_{1 \perp} \mathrm{d} \omega_1}{E}
  \cdot   \frac{\mathrm{d}^2 q_{2 \perp} \mathrm{d} \omega_2}{xE}.
  \label{phase-transition}
\end{equation}

The photon density matrix for the photon emitted by the proton is
\begin{equation}
  \rho_1^{\mu \nu}
  = -\frac{1}{2 q_1^2}
     \Tr \left\{
       (\hat p'_1 + m_p)
       \left(
           F_1(Q_1^2) \gamma^\mu
         + F_2(Q_1^2) \frac{\sigma^{\mu \alpha} q_1^\alpha}{2 m_p}
       \right)
       (\hat p_1 + m_p)
       \left(
           F_1(Q_1^2) \gamma^\nu
         - F_2(Q_1^2) \frac{\sigma^{\nu \alpha} q_1^\alpha}{2 m_p}
       \right)
     \right\},
  \label{rho-form-factors}
\end{equation}
where $F_1(Q_1^2)$ and $F_2(Q_1^2)$ are the Dirac and Pauli form factors. The
Sachs form factors~\eqref{sachs-form-factors} are their linear combinations:
\begin{equation}
  \begin{aligned}
    G_E(Q_1^2) &= F_1(Q_1^2) - \frac{Q_1^2}{4 m_p^2} F_2(Q_1^2), \\
    G_M(Q_1^2) &= F_1(Q_1^2) + F_2(Q_1^2).
  \end{aligned}
\end{equation}
Eq.~\eqref{rho-form-factors} simplifies to
\begin{equation}
  \rho_1^{\mu \nu}
  = -\left( g^{\mu \nu} - \frac{q_1^\mu q_1^\nu}{q_1^2} \right) G_M^2(Q_1^2)
  - \frac{(2 p_1 - q_1)^\mu (2 p_1 - q_1)^\nu}{q_1^2} D(Q_1^2),
  \label{rho1}
\end{equation}
where $D(Q_1^2)$ is defined in Eq.~\eqref{form-factor}.

The photon density matrix for the photon emitted by the quark is
\begin{equation}
  \rho_2^{\mu \nu}
  = -\frac{1}{2 q_2^2}
     \Tr \left\{ \hat p'_2 \gamma^\mu \hat p_2 \gamma^\nu \right\}
  = -\left( g^{\mu \nu} - \frac{q_2^\mu q_2^\nu}{q_2^2} \right)
  - \frac{(2 p_2 - q_2)^\mu (2 p_2 - q_2)^\nu}{q_2^2}.
  \label{rho2}
\end{equation}

It is convenient to consider the lepton pair production in the basis of virtual
photon helicity states. In the center of mass system of the colliding photons,
let $q_1 = (\tilde \omega_1, 0, 0, \tilde q)$, $q_2 = (\tilde \omega_2, 0, 0,
-\tilde q)$. The standard set of orthonormal 4-vectors orthogonal to $q_1$ and
$q_2$ is
\begin{equation}
  \begin{aligned}
       e_1^+ &= \frac{1}{\sqrt{2}} (0, -1, -i, 0),
    &  e_1^- &= \frac{1}{\sqrt{2}} (0,  1, -i, 0),
    &  e_1^0 &= \frac{i}{\sqrt{-q_1^2}} (\tilde q, 0, 0, \tilde \omega_1),
    \\ e_2^+ &= \frac{1}{\sqrt{2}} (0,  1, -i, 0),
    &  e_2^- &= \frac{1}{\sqrt{2}} (0, -1, -i, 0),
    &  e_2^0 &= \frac{i}{\sqrt{-q_2^2}} (-\tilde q, 0, 0, \tilde \omega_2).
  \end{aligned}
\end{equation}
Due to the conservation of vector current, the covariant density matrices
$\rho_i^{\mu \nu}$ satisfy $q_1^\mu \rho_1^{\mu \nu} = q_2^\mu \rho_2^{\mu \nu}
= 0$. Thus, we can write
\begin{align}
  \rho_i^{\mu \nu}
  &= \sum\limits_{a,b} (e_i^{a \mu})^* e_i^{b \nu} \rho_i^{ab},
  \\
  \rho_i^{ab} &= (-1)^{a+b} e_i^{a \mu} (e_i^{b \nu})^* \rho_i^{\mu \nu},
  \label{density-helicity}
\end{align}
where $a, b \in \{ \pm 1, 0 \}$, and $\rho_i^{ab}$ are the density matrices in
the helicity representation. The amplitudes of the lepton pair production in the
helicity basis $M_{ab}$ comply to the equation
\begin{equation}
  \rho_1^{\mu \nu} \rho_2^{\alpha \beta} M_{\mu \alpha} M^*_{\nu \beta}
  = (-1)^{a + b + c + d}
    \rho_1^{ab} \rho_2^{cd} M_{ac} M^*_{bd}.
\end{equation}
In this expression, non-diagonal terms (those with $a \ne b$ or $c \ne d$)
originate from the interference and cancel out when integrated over azimuthal
angles of the proton and the quark in the final state~\cite{prep15-181,
zpc52-427}.  Contributions to the cross section for lepton pair production by
longitudinally polarized photons are proportional to $Q_1^2 / W^2 \le \hat q^2 /
W^2 \ll 1$ and $Q_2^2 / W^2 \le (p_T^{\ell \ell})^2 / W^2 \ll 1$ and are
neglected in the following. Thus, we can rewrite~\eqref{master} in the helicity
representation:
\begin{multline}
  \mathrm{d} \sigma(p q \to p + \ell^+ \ell^- + q)
  = Q_q^2 (4 \pi \alpha)^2
    \left(
        \rho_1^{++} \rho_2^{++} \abs{M_{++}}^2
      + \rho_1^{++} \rho_2^{--} \abs{M_{+-}}^2
      + \rho_1^{--} \rho_2^{++} \abs{M_{-+}}^2
    + {} \right. \\ \left.
      \rho_1^{--} \rho_2^{--} \abs{M_{--}}^2
    \right)
    \frac{(2 \pi)^4 \delta^{(4)}(q_1 + q_2 - k_1 - k_2) \mathrm{d} \Gamma}
         {4 p_1 p_2}
    \cdot \frac{\mathrm{d}^2 q_{1 \perp} \mathrm{d} \omega_1}{(2 \pi)^3 q_1^2 E}
    \cdot \frac{\mathrm{d}^2 q_{2 \perp} \mathrm{d} \omega_2}{(2 \pi)^3 q_2^2 xE}
    \cdot f_q(x, Q_2^2) \mathrm{d} x.
  \label{xsection-helicity}
\end{multline}

Substitution of~\eqref{rho1}, \eqref{rho2} to~\eqref{density-helicity} yields
the following expressions
\begin{equation}
  \begin{aligned}
    \rho_1^{++} = \rho_1^{--}
    &= G_M^2(Q_1^2) + \frac{2 p_{1 \perp}^2}{q_1^2} D(Q_1^2), \\
    \rho_2^{++} = \rho_2^{--}
    &= 1 + \frac{2 p_{2 \perp}^2}{q_2^2},
  \end{aligned}
\end{equation}
where $p_{i \perp}$ is the component of the momentum $p_i$ orthogonal to $q_i$
in the c.m.s. of the colliding photons. To calculate it, we
follow~\cite{prep15-181} and introduce the symmetrical tensor
\begin{equation}
  R^{\mu \nu}(q_1, q_2)
  = -g^{\mu \nu}
  + \frac{
        q_1 q_2 \cdot (q_1^\mu q_2^\nu + q_1^\nu q_2^\mu)
      - q_1^2 q_2^\mu q_2^\nu
      - q_2^2 q_1^\mu q_1^\nu
    }{ (q_1 q_2)^2 - q_1^2 q_2^2 }.
\end{equation}
This tensor is the metric tensor of the subspace orthogonal to $q_1$ and $q_2$,
and has the following properties:
\begin{equation}
    q_i^\mu R^{\mu \nu} = 0,
  ~ R^{\alpha \beta} R^{\beta \gamma} = -R^{\alpha \gamma},
  ~ R^{\alpha \beta} R^{\alpha \beta} = 2.
\end{equation}
Then $p_{i, \perp}^\mu = -R^{\mu \nu} p_i^{\nu}$, and\footnote{
  The identity $q_1^2 = 2 p_1 q_1$ helps in deriving Eq.~\eqref{rho}.
}
\begin{equation}
  \begin{aligned}
       \rho_1^{++}
     = \rho_1^{--}
    &= G_M^2(Q_1^2)
     + D(Q_1^2) \left[
         \frac{2 m_p^2}{q_1^2}
         + \frac12 \left(
             \frac{(2 p_1 q_2 - q_1 q_2)^2}{(q_1 q_2)^2 - q_1^2 q_2^2} - 1
           \right)
       \right],
    \\
       \rho_2^{++}
     = \rho_2^{--}
    &= \frac12
     + \frac{2 m_q^2}{q_2^2}
     + \frac12 \cdot \frac{(2 p_2 q_1 - q_1 q_2)^2}{(q_1 q_2)^2 - q_1^2 q_2^2}.
  \end{aligned}
  \label{rho}
\end{equation}
With the help of equations~\eqref{p1-p2}--\eqref{virtuality2}, under the
assumptions $q_1^2 q_2^2 \ll (q_1 q_2)^2$, $\omega_1 \ll E$, $\omega_2 \ll x E$,
these expressions simplify to
\begin{equation}
  \begin{aligned}
     \rho_1^{++} = \rho_1^{--}
    &\approx G_M^2(Q_1^2)
      + 2 D(Q_1^2) \left[
         \frac{m_p^2}{q_1^2} + \left( \frac{E}{\omega_1} \right)^2
      \right]
     \approx D(Q_1^2) \cdot \frac{2 E^2 q_{1 \perp}^2}{\omega_1^2 Q_1^2},
  \\
     \rho_2^{++} = \rho_2^{--}
    &\approx 1
     + 2 \left[
         \frac{m_q^2}{q_2^2} + \left( \frac{x E}{\omega_2} \right)^2
     \right]
     \approx \frac{2 x^2 E^2 q_{2 \perp}^2}{\omega_2^2 Q_2^2}.
  \end{aligned}
  \label{rho-approx}
\end{equation}

Substitution of equations \eqref{rho-approx} to~\eqref{xsection-helicity} yields
\begin{multline}
  \mathrm{d} \sigma(pq \to p + \ell^+ \ell^- + q)
  \approx
  \left( \frac{2 Q_q \alpha}{\pi} \right)^2
  \frac{q_1 q_2}{p_1 p_2}
  x E^2
  \sigma(\gamma \gamma \to \ell^+ \ell^-)
  D(Q_1^2)
  \frac{q_{1 \perp}^3 \mathrm{d} q_{1 \perp}}{Q_1^4}
  \frac{\mathrm{d} \omega_1}{\omega_1^2}
  \times \\
  \frac{q_{2 \perp}^3 \mathrm{d} q_{2 \perp}}{Q_2^4}
  \frac{\mathrm{d} \omega_2}{\omega_2^2}
  \cdot f_q(x, Q_2^2) \mathrm{d} x,
\end{multline}
where
\begin{equation}
  \sigma(\gamma \gamma \to \ell^+ \ell^-)
  = \int
      \frac14 [
        \abs{M_{++}}^2 + \abs{M_{+-}}^2 + \abs{M_{-+}}^2 + \abs{M_{--}}^2
      ]
      \frac{(2 \pi)^4 \delta^{(4)}(q_1 + q_2 - k_1 - k_2) \mathrm{d} \Gamma}
           {4 q_1 q_2},
\end{equation}
is the cross for lepton pair production in a collision of two real unpolarized
photons. Integration over $q_{1 \perp}$ yields the equivalent photon spectrum of
proton~\eqref{proton-spectrum}. To derive the fiducial cross section, we change
the integration variables: $\mathrm{d} \omega_1 \mathrm{d} \omega_2 =
\frac{W}{2} \mathrm{d} W \mathrm{d} y$. Then
\begin{equation}
  \mathrm{d} \sigma(pq \to p + \ell^+ \ell^- + q)
  \approx
  \frac{2 Q_q^2 \alpha}{\pi}
  \; n_p \left( \frac{W}{2} \mathrm{e}^y \right)
  \, \sigma(\gamma \gamma \to \ell^+ \ell^-)
  \; \mathrm{e}^y
  \; \frac{q_{2 \perp}^3 \mathrm{d} q_{2 \perp}}{Q_2^4}
  \mathrm{d} W
  \mathrm{d} y
  \; f_q(x, Q_2^2) \mathrm{d} x.
\end{equation}

Introducing the function
\begin{equation}
  n_q(\omega)
  = \frac{2 Q_q^2 \alpha}{\pi \omega}
    \int\limits_{\omega / E}^1 \mathrm{d} x
    \int\limits_0^{p_T^{\ell \ell}} \mathrm{d} q_{2 \perp}
    \, \frac{q_{2 \perp}^3}{Q_2^4} f_q(x, Q_2^2),
\end{equation}
which can be loosely interpreted as the equivalent photon spectrum of quark $q$
(cf.~\eqref{proton-spectrum}), we derive the equation for the fiducial cross
section similar to~\eqref{xsection}, \eqref{luminosity}:
\begin{multline}
  \frac{\mathrm{d} \sigma_\text{fid.}(pp \to p + \ell^+ \ell^- + X)}
       {\mathrm{d} W}
  = \sum\limits_q
    \int\limits_{\max \left( \hat p_T, \frac{W}{2 \cosh \hat \eta} \right)}^
                {W / 2}
    \mathrm{d} p_T
    \frac{\mathrm{d} \sigma(\gamma \gamma \to \ell^+ \ell^-)}{\mathrm{d} p_T}
    \times \\
    \frac{W}{2}
    \int\limits_{\max(-\hat y, \tilde y)}^{\min(\hat y, \tilde Y)}
    \mathrm{d} y
    \, n_p \left( \frac{W}{2} e^y    \right)
    \, n_q \left( \frac{W}{2} e^{-y} \right).
\end{multline}
Similar to~\eqref{double-counting}, to calculate the cross section measured
in~\cite{2009.14537}, this value should be multiplied by 2 to take into account
that either of the protons can hit the detector:
\begin{equation}
  \sigma_\text{fid, \cite{2009.14537}}(pp \to p + \ell^+ \ell^- + X)
  = 2 \sigma_\text{fid.}(pp \to p + \ell^+ \ell^- + X).
\end{equation}

Using the parton distribution functions
\texttt{MSHT20nnlo\_as118}~\cite{2012.04684} provided by the
LHAPDF~\cite{1412.7420} library, we get the cross sections
\begin{align}
  \sigma_\text{fid, \cite{2009.14537}}(pp \to p + \mu^+ \mu^- + X)
  &= 9.6~\text{fb},
  \label{inclusive-mu}
  \\
  \sigma_\text{fid, \cite{2009.14537}}(pp \to p + e^+ e^- + X)
  &= 11.4~\text{fb}.
  \label{inclusive-e}
\end{align}

In order to estimate the accuracy of these cross sections, we have calculated
them with the shifted arguments of parton density functions:
\begin{align}
  f_q(x, Q_2^2/2): \quad
  &
  \left\{
    \begin{aligned}
      \sigma(pp \to p + \mu^+ \mu^- + X) &= 7.7~\text{fb}, \\
      \sigma(pp \to p + e^+ e^- + X) &= 9.1~\text{fb}, \\
    \end{aligned}
  \right.
  \label{inclusive-uncertainty-mu}
  \\
  f_q(x, 2 Q_2^2): \quad
  &
  \left\{
    \begin{aligned}
      \sigma(pp \to p + \mu^+ \mu^- + X) &= 11.5~\text{fb}, \\
      \sigma(pp \to p + e^+ e^- + X) &= 13.6~\text{fb}. \\
    \end{aligned}
  \right.
  \label{inclusive-uncertainty-e}
\end{align}

We have checked that the contribution from the region of small $Q^2$ where
the PDFs are known with less accuracy is small. To do that we have calculated
the inelastic contribution with much stronger cut on the transverse momentum:
$Q^2 \approx q_\perp^2 < 1~\text{GeV}^2$. We get $2.2$~fb for the production of
muons and $2.6$~fb for the production of electrons. Therefore, it is about 20\%
of the inelastic contribution, and of the order of what we have estimated for
the PDFs uncertainty. We add this as a separate source of the uncertainty of the
inelastic contribution.

We must also stress here that PDFs alone do not describe all inelastic
contribution and there is a contribution to proton structure functions
from resonance phenomena and other effects, see
\cite{1708.01256,2107.02535,1510.00294}. From Figure~18 of
Ref.~\cite{1708.01256} and Table~I of Ref.~\cite{2107.02535} we can
see that the contribution of these effects is non-vanishing but at the
level of 10--15\% from elastic contribution. We do not aim at such
level of precision and take into account just major contributions.

\section{Conclusions}

\label{s:conclusion}

Let us compare our results with experimental data from~\cite{2009.14537}:
\begin{align}
  \sigma_{\mu \mu + p}^\text{exp.}
  &= 7.2
  \pm 1.6~\text{(stat.)}
  \pm 0.9~\text{(syst.)}
  \pm 0.2~\text{(lumi.)}
  ~\text{fb},
  \\
  \sigma_{ee + p}^\text{exp.}
  &= 11.0
  \pm 2.6~\text{(stat.)}
  \pm 1.2~\text{(syst.)}
  \pm 0.3~\text{(lumi.)}
  ~\text{fb}.
\end{align}
Summing up the cross sections calculated in Sections~\ref{s:exclusive}
and~\ref{s:inclusive}, we get:
\begin{align}
  \tilde \sigma_{\mu \mu + p}^\text{theor.} &= 18 \pm 3~\text{fb}, \\
  \tilde \sigma_{e e + p}^\text{theor.} &= 22 \pm 3~\text{fb},
\end{align}
where the uncertainty values were obtained by comparing
Eqs.~\eqref{inclusive-uncertainty-mu}, \eqref{inclusive-uncertainty-e}
and~\eqref{inclusive-mu}, \eqref{inclusive-e} and include the uncertainty of the
contribution from the region of small $Q^2$, see the paragraph
after~\eqref{inclusive-uncertainty-e}. Also was taken into account the very small
contribution from quark mass uncertainty in~\eqref{q2}, see
footnote~\ref{fn:quark-mass} at page~\pageref{fn:quark-mass}.

The so-called survival factor takes into account the diminishing of
the cross sections due to breaking of both protons occurring when the
protons scatter with small impact parameter $b$. The survival factor
decreases when the invariant mass of the lepton pair grows and for the
elastic cross section at $W \sim 100$~GeV it approximately equals to
$0.9$ according to Fig.~3 from~\cite{2106.14842}.

Table~1 in~\cite{2009.14537} contains results for the cross section obtained by
Monte Carlo simulation. From the first two lines of this Table it follows that
the survival factor (the probability for the colliding protons to avoid strong
interactions) decreases the cross section by approximately factor $1.5$. Here
the values of cross sections for $S_\text{surv} = 1$ are the combined results of
LPAIR and HERWIG event generators, and the results for the survival factor from
papers~\cite{1410.2983, 1508.02718} are used. Results of SUPERCHIC4
code~\cite{2007.12704} which takes the survival factor into account are within
one standard deviation, see the third line in Table~1.

In the case of the elastic process, according to~\cite{2106.14842} the cross
section is approximately 10\% less, while in the case of elastic-inelastic
scattering the cross section is approximately 50\% less, see the lower left
panel of Fig.~28 from Ref.~\cite{2107.02535}.

We see that when the survival factor is taken into account the derived formulae
are in agreement with experimental data at the level of 2--3 standard
deviations. In the case of elastic process, the survival factor is calculated
in~\cite{2106.14842} and can be easily taken into account without resorting to
Monte Carlo simulations. Whether calculations of survival factor for
semiexclusive process can be performed in the similar way is an interesting
subject for further study.

Monte Carlo codes for photon-photon processes in UPC of protons and/or nuclei are
presented in the recent paper~\cite{2207.03012}.

Our calculations were performed with the help of \texttt{libepa}~\cite{libepa}.

We are grateful to the referee for emphasizing the necessity of
analyzing the contribution of low $Q^2$ domain to the inelastic cross
section.

The authors were supported by the Russian Science Foundation grant No
19-12-00123-$\Pi$.

\newcommand{\arxiv}[1]{arXiv:\nolinebreak[3]\href{http://arxiv.org/abs/#1}{#1}}

\end{document}

%% file: tikz.tex
\usepackage{tikz}
\usetikzlibrary{arrows}
\usetikzlibrary{decorations.pathmorphing}
\usetikzlibrary{decorations.markings}
\usetikzlibrary{calc}

\tikzset{
  >=stealth',
  midarrow/.style={
    postaction={
      decorate,
      decoration={markings, mark=at position .55 with {\arrow{>}}}
    }
  },
  fermion/.style=midarrow,
  photon/.style={
    decorate,
    decoration={snake, amplitude=2pt, segment length=8pt}
  },
  boson/.style={
    decorate,
    decoration={snake, amplitude=2pt, segment length=8pt}
  },
  gluon/.style={
    decorate,
    decoration={coil, amplitude=4pt, segment length=5pt}
  },
  scalar/.style=densely dashed,
  arrowsnake/.style={
    preaction={photon, draw},
    postaction=midarrow
  }
}